\pdfoutput=1  
\documentclass{article}
\usepackage{graphicx}
\usepackage{amsfonts}
\usepackage{amssymb}
\usepackage{url}

\def\e{{\varepsilon}}

\def\s{{\sigma}}
\def\hull{\mathop{\rm hull}\nolimits}
\def\sp{\mathop{\rm sp}\nolimits}

\def\st{{\sc Stretch}}
\def\hp{{\sc Hop}}
\def\sw{{\sc Swap}}
\def\tw{{\sc Twang}}
\def\twc{{\sc TwangCascade}}

\def\fm{{\sc ForwardMove}}

\newcommand{\hide}[1]{}

\newcommand\Pw{\mathcal{P}}
\newcommand{\squeezelist}{\setlength{\itemsep}{0pt}}

\newtheorem{theorem}{{\bf Theorem}}
\newtheorem{corollary}[theorem]{Corollary}
\newtheorem{lemma}[theorem]{Lemma}

\newtheorem{definition}[theorem]{Definition}

\newcommand{\lemlab}[1]{\label{lemma:#1}}

\newcommand{\corlab}[1]{\label{cor:#1}}

\newcommand{\figlab}[1]{\label{fig:#1}}
\newcommand{\seclab}[1]{\label{sec:#1}}

\newcommand{\lemref}[1]{\ref{lemma:#1}}

\newcommand{\corref}[1]{\ref{cor:#1}}

\newcommand{\secref}[1]{\ref{sec:#1}}
\newcommand{\figref}[1]{\ref{fig:#1}}

\newcommand{\ABox}{
\raisebox{3pt}{\framebox[6pt]{\rule{6pt}{0pt}}}
}
\newenvironment{pf}{{\bf Proof:}}{\hfill\ABox}

{\makeatletter
 \gdef\xxxmark{%
   \expandafter\ifx\csname @mpargs\endcsname\relax 
     \expandafter\ifx\csname @captype\endcsname\relax 
       \marginpar{xxx}
     \else
       xxx 
     \fi
   \else
     xxx 
   \fi}
 \gdef\xxx{\@ifnextchar[\xxx@lab\xxx@nolab}
 \long\gdef\xxx@lab[#1]#2{{\bf [\xxxmark #2 ---{\sc #1}]}}
 \long\gdef\xxx@nolab#1{{\bf [\xxxmark #1]}}
 \gdef\turnoffxxx{\long\gdef\xxx@lab[##1]##2{}\long\gdef\xxx@nolab##1{}}%
}

\title{Connecting Polygonizations via Stretches and Twangs}
\author{Mirela Damian%
   \thanks{Dept. of Computer Science, Villanova Univ., Villanova,
    PA 19085, USA.
   \protect\url{mirela.damian@villanova.edu}.}
\and
Robin Flatland%
   \thanks{Dept. of Computer Science, Siena College, Loudonville, NY 12211, USA.
    \protect\url{flatland@siena.edu}.}
\and
Joseph O'Rourke%
    \thanks{Dept. of Computer Science, Smith College, Northampton, MA
      01063, USA.
      \protect\url{orourke@cs.smith.edu}.
       }
\and
Suneeta Ramaswami%
    \thanks{Dept. of Computer Science, Rutgers University,
       Camden, NJ 08102, USA.
   \protect\url{rsuneeta@camden.rutgers.edu}.}
}
\date{}

\begin{document}
\maketitle

\begin{abstract}
\noindent We show that the space of polygonizations of a fixed
planar point set $S$ of $n$ points is connected by
$O(n^2)$ ``moves'' between simple polygons.
Each move is composed of a sequence of
atomic moves called ``stretches'' and ``twangs''.
These atomic moves walk between weakly simple
``polygonal wraps'' of $S$.
These moves show promise to serve as a basis for generating random polygons.
\end{abstract}

\maketitle

\section{Introduction}
\seclab{Introduction} This paper studies polygonizations of a fixed
planar point set $S$ of $n$ points. Let the $n$ points be labeled
$p_i$, $i=0,1,\ldots,n{-}1$. A \emph{polygonization} of $S$ is a
permutation $\s$ of $\{0,1,\ldots,n{-}1\}$ that determines a polygon:
$P = P_{\s}  = (p_{\s(0)}, \ldots, p_{\s(n{-}1)})$ is a simple
(non-self-intersecting) polygon. We will abbreviate ``simple
polygon'' to \emph{polygon} throughout.
As long as $S$ does not lie in one line, which we will henceforth assume,
there is at least one
polygon whose vertex set is $S$. A point set $S$ may have as few as
$1$ polygonization, if $S$ is in convex position,\footnote{
 $S$ is in convex position if every point in $S$ is on the hull of $S$.
}
and as many as
$2^{\Theta(n)}$ polygonizations. For the latter, see
Fig.~\figref{one.pocket.polygonization}a.

Our goal in this work is to develop a computationally natural and
efficient method to explore all polygonizations of a fixed set $S$.
One motivation is the generation of ``random polygons'' by first
generating a random $S$ and then selecting a random polygonization
of $S$. Generating random polygons efficiently is a long unsolved
problem; only heuristics~\cite{ah-hgrp-96} or algorithms for special
cases~\cite{zssm-grpgv-96} are known. Our work can be viewed as
following a suggestion 
 in the latter paper:
\begin{quotation}
\noindent ``start with a ... simple polygon and apply some
simplicity-preserving, reversible operations ... with the property
that any simple polygon is reachable by a sequence of operations''
\end{quotation}

\noindent Our two operations are called \emph{stretch} and
\emph{twang} (defined in Section~\secref{Stretches.Twangs}). Neither
is simplicity preserving, but they are nearly so in that they
produce polygonal wraps defined as follows.

\begin{definition}
\emph{A} polygonal wrap \emph{$\Pw_\sigma$ is determined by
a sequence $\s$ of point indices drawn from $\{0,1,\ldots,n{-}1\}$
with the following properties:
\begin{enumerate}
\squeezelist
\item Every index in $\{0,1,\ldots,n{-}1\}$ occurs in $\s$.
\item Indices may be repeated.  If index $i$ appears
at least twice in $\s$, we call $p_i$ a \emph{point of double contact}.
\item For sufficiently small $\e > 0$, there exists a perturbation
within an $\e$-disk of the points in double contact, separating each
such point into two or more points, so that there is a simple closed curve
$C$ that passes through the perturbed points in $\s$ order.
Sometimes such a $P$ is called ``weakly simple'' because its
violations of simplicity (i.e., its self-touchings) avoid proper
crossings.\footnote{
   Two segments properly cross if they share a point $x$ in the relative interior of
   both, and cross transversely at $x$.}
\end{enumerate}
}
\label{def:polygonal.wrap}
\end{definition}

\vspace{-0.5em} 
\noindent Fig.~\figref{one.pocket.polygonization}b
shows a polygonal wrap 
with five double-contacts
($p_1,p_4,p_5,p_8$ and $p_9$). Note that a polygon is a polygonal
wrap without double-contact
points. 

Stretches and twangs take one polygonal wrap to another. A stretch
followed by a natural sequence of twangs, which we call a
\emph{cascade}, constitutes a \emph{forward move}. Forward moves
(described in further detail in Section~\secref{Twang.Cascades})  
take a polygon to a polygon, i.e., they are simplicity preserving.
Reverse moves will be introduced in Section~\secref{Reverse}. A
\emph{move} is either a forward or a reverse move. We call a stretch
or twang an \emph{atomic move} to distinguish it from the more
complex forward and reverse moves.

Our main result is that the configuration space of polygonizations for
a fixed $S$ is connected by forward/reverse moves, each of which is
composed of a number of stretches and twangs, and that the diameter of
the space is $O(n^2)$ moves.
We can bound the worst-case number of atomic moves constituting a
particular forward/reverse move
by the geometry of the point set.
Experimental results on random point sets show that,
in the practical situation that is one of our motivations, the bound
is small, perhaps even constant.
We have also established loose bounds on the worst-case number of atomic
operations as a function of $n$:
an exponential upper bound and a quadratic lower bound. 
Tightening these bounds has so far proven elusive
and is an open problem.

One can view our work as in the
tradition of connecting discrete structures (e.g., triangulations,
matchings) via local moves (e.g., edge flips, edge swaps).  Our result
is comparable to that in~\cite{ls-utstp-82}, which shows connectivity
of polygonizations in $O(n^3)$ edge-edge swap moves through
intermediate self-crossing polygons. The main novelty of our work is
that the moves, and even the stretches and twangs, never lead to
proper crossings, for polygonal wraps have no such crossings. We explore
the possible application to random polygons briefly in
Section~\secref{Random.Polygons}. For the majority of this paper, we
concentrate on defining the moves and establishing connectivity.

\begin{figure}[htbp]
\centering
\includegraphics[width=0.9\linewidth]{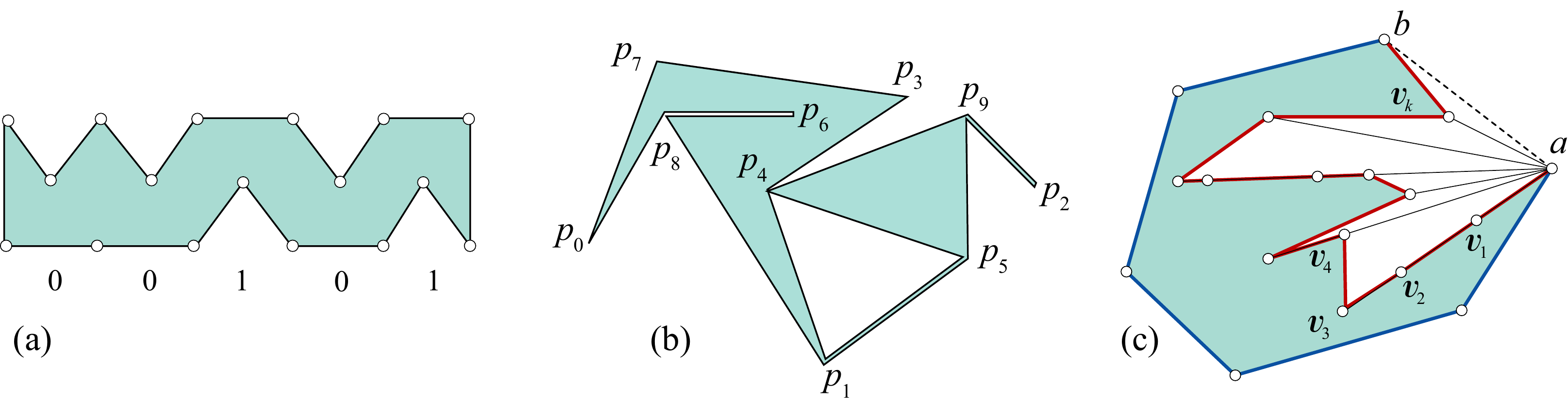}
\vspace{-0.5em}\caption{Examples. (a) A set of $n=3k+2$ points that
admits $2^k$ polygonizations. (b) Polygonal wrap $\Pw_\sigma$ with
$\sigma=(0,8,6,8,1,5,9,2,9,4,5,1,4,3,7)$ (c) A polygonization with
one pocket with lid $ab$.} \figlab{one.pocket.polygonization}
\end{figure}

We begin by defining pockets, which play a central role in our
algorithms for polygonal transformations. Then in
Section~\secref{Swaps.Hops} we describe two natural operations that
transform one polygon into another but fail to achieve connectivity of
the configuration space of polygonizations, which motivates our
definitions of stretches and twangs in
Section~\secref{Stretches.Twangs}. Following these preliminaries, we
establish connectivity and compute the diameter in
Sections~\secref{single.pocket.reduction}--\secref{Connectivity}.
We conclude with open problems in Section~\secref{Open.Problems}.

\subsection{Pockets and Canonical Polygonization}
\seclab{Pockets} Let $P$ be a polygonization of $S$. A hull edge
$ab$ that is not on $\partial P$ is called a \emph{pocket lid}. The
polygon external to $P$ bounded by $P$ and $ab$ is a \emph{pocket}
of $P$.

\begin{lemma}
Any point set $S$ not in convex position has a polygonization with
one pocket only~\emph{\cite{chuz-peps-92}}. \lemlab{one.pocket}
\end{lemma}
For a fixed hull edge $ab$, we define the canonical polygonization
of $S$ to be a polygon with a single pocket with lid $ab$ (cf.
Lemma~\lemref{one.pocket}) in which the pocket vertices are ordered
by angle about vertex $a$, and from closest to farthest from $a$ if
along the same line through $a$. We call this ordering the
\emph{canonical order} of the pocket vertices; see
Fig.~\figref{one.pocket.polygonization}c.

\vspace{-1em}
\section{Polygonal Transformations}

Let $P$ be a polygon defined by a circular index sequence $\sigma$.
We examine operations that permute this sequence, transforming $P$
into a new polygon with the same set of vertices
linked in a different order. 
Throughout the paper we use $\triangle abc$ to denote the
closed triangle with corners
$a$, $b$ and $c$.

\subsection{Swaps and Hops}
\seclab{Swaps.Hops} We begin by defining two natural transformation
operations, a \emph{swap} and a \emph{hop}. A swap operation
is a transposition of
two consecutive vertices of $P$ that results in a new
(non-self-intersecting) polygon. Fig.~\figref{swaps}a illustrates
the swap operation. It is well known that transpositions connect all
permutations, but this is not the case for swaps. Because
we require that the resulting polygon be simple, a vertex pair
cannot be swapped if the operation results in a self-intersecting
polygon. Fig.~\figref{swaps}b shows an example of a polygon for
which no vertex pair can be swapped without creating an edge
crossing. Thus, swaps do not suffice to connect all polygonizations,
which motivates our definition of a more powerful move,
which we call a \emph{hop}.

\begin{figure}[htbp]
\centering
\begin{tabular}{cc@{\hspace{0.05\linewidth}}cc}
\raisebox{0.2in}{(a)} &
\raisebox{0.1in}{\includegraphics[width=0.5\linewidth]{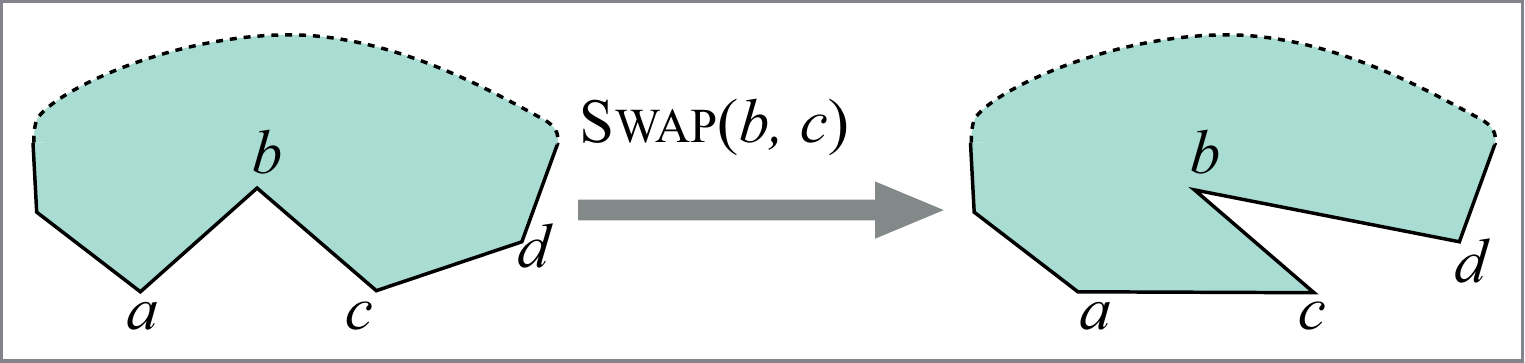}}
& \raisebox{0.2in}{(b)} &
\includegraphics[width=0.3\linewidth]{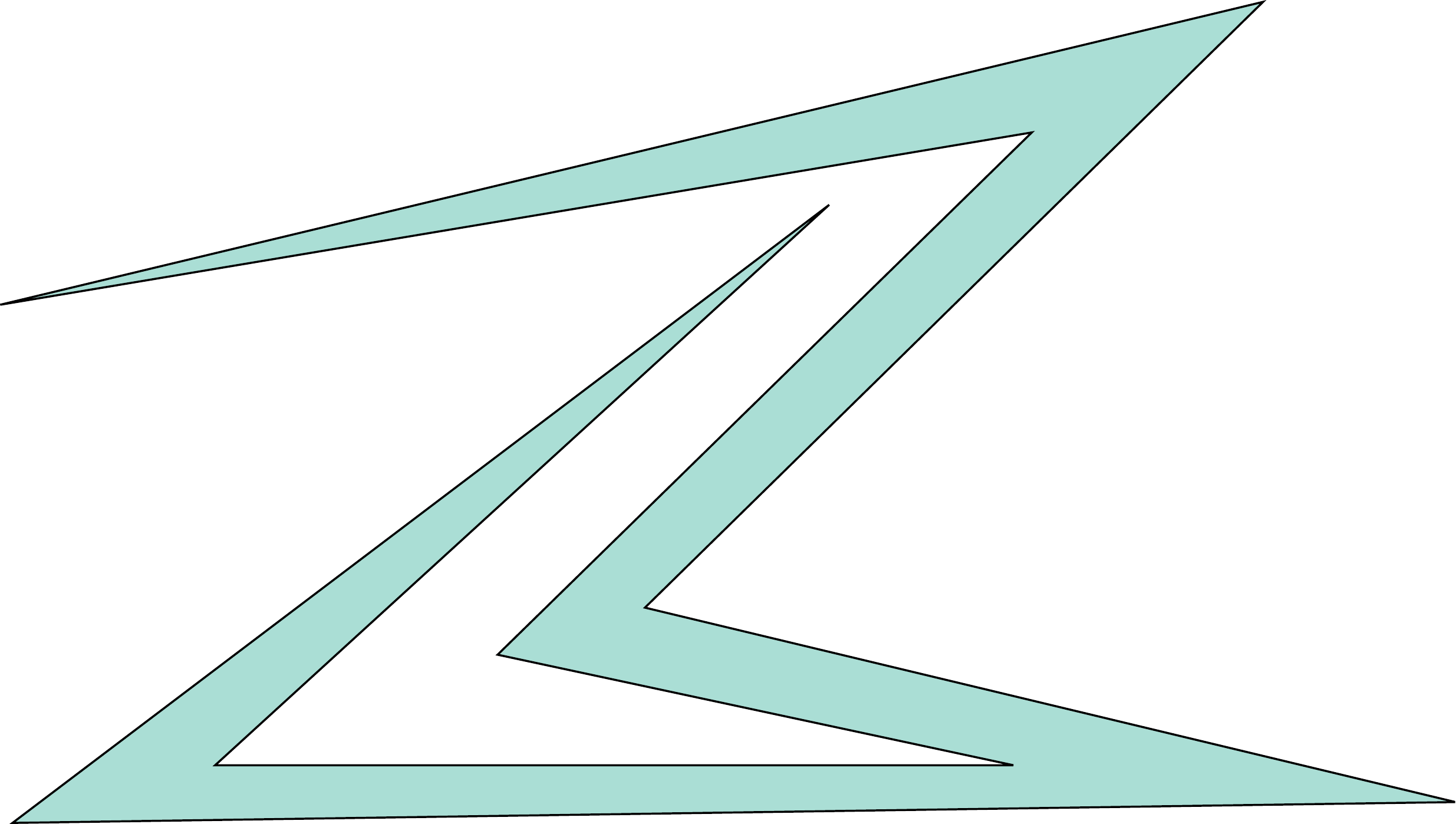}
\end{tabular}
\vspace{-0.5em} \caption{(a) \sw$(b,c)$ illustrated (b) Polygon
admitting no swaps.} \figlab{swaps}
\end{figure}

The hop operation generalizes the swap by allowing a vertex to hop
to any position in the permutation, as long as the resulting polygon
is simple. Fig.~\figref{hops}a shows the stretching of the edge $ab$ down to
vertex $v$, effectively ``hopping'' $v$ between
$a$ and $b$ in the permutation. We denote this operation
by \hp$(e, v)$, where $e = ab$ (note the first argument
is \emph{from} and the second \emph{to}).

To specify the conditions under which a hop operation is valid, we
introduce some definitions, which will be used subsequently as well.
A polygon $P$ has two sides, the interior of $P$ and the
exterior of $P$. Let $abc=(a,b,c)$ be three vertices consecutive in
the polygonization $P$. For noncollinear vertices, we distinguish
between the \emph{convex side} of $b$, that side of $P$
with angle $\angle abc$ smaller than $\pi$, and the \emph{reflex
side} of $b$, the side of $P$ with angle $\angle abc$
larger than $\pi$.
Note that this definition ignores which side is the interior and
which side is the exterior of $P$, and so is unrelated to whether
$b$ is a convex or a reflex vertex in $P$. Every true vertex has a
convex and a reflex side (collinear vertices will be discussed in
Section~\secref{Stretches.Twangs}).
To ensure that the resulting polygon is simple, \hp$(e, v)$ is
valid iff the following two conditions hold: (1) the triangle
induced by the two edges incident to $v$ is empty of other polygon
vertices and (2) the triangle induced by $e$ and $v$ lies on the
reflex side of $v$ and is empty of other polygon vertices.

Although more powerful than a swap, there also exist polygons that do
not admit any hops. Fig.~\figref{hops}b shows an example (the
smallest we could find) in which each edge-vertex pair violates one
or both of the two conditions above. This example shows that hops do
not suffice to connect all polygonizations. 

\vspace{-0.5em}
\begin{figure}[htbp]
\centering
\begin{tabular}{cc@{\hspace{0.1\linewidth}}cc}
\raisebox{0.2in}{(a)} &
\raisebox{0.1in}{\includegraphics[width=0.52\linewidth]{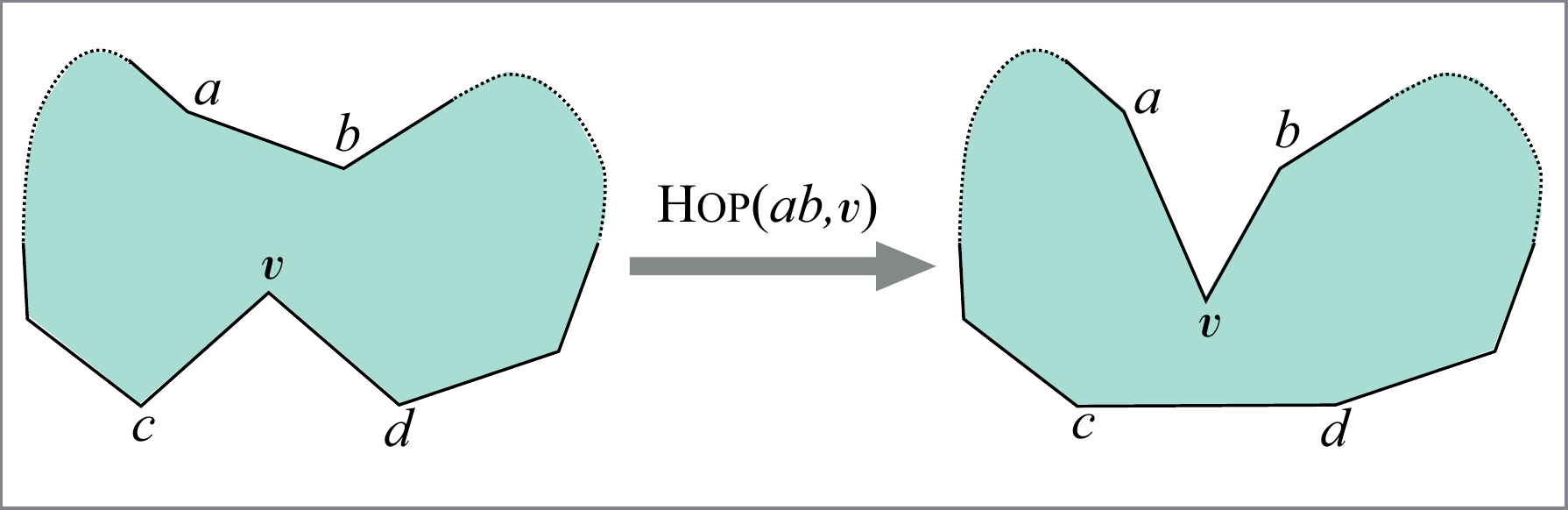}}
& \raisebox{0.2in}{(b)} &
\includegraphics[width=0.25\linewidth]{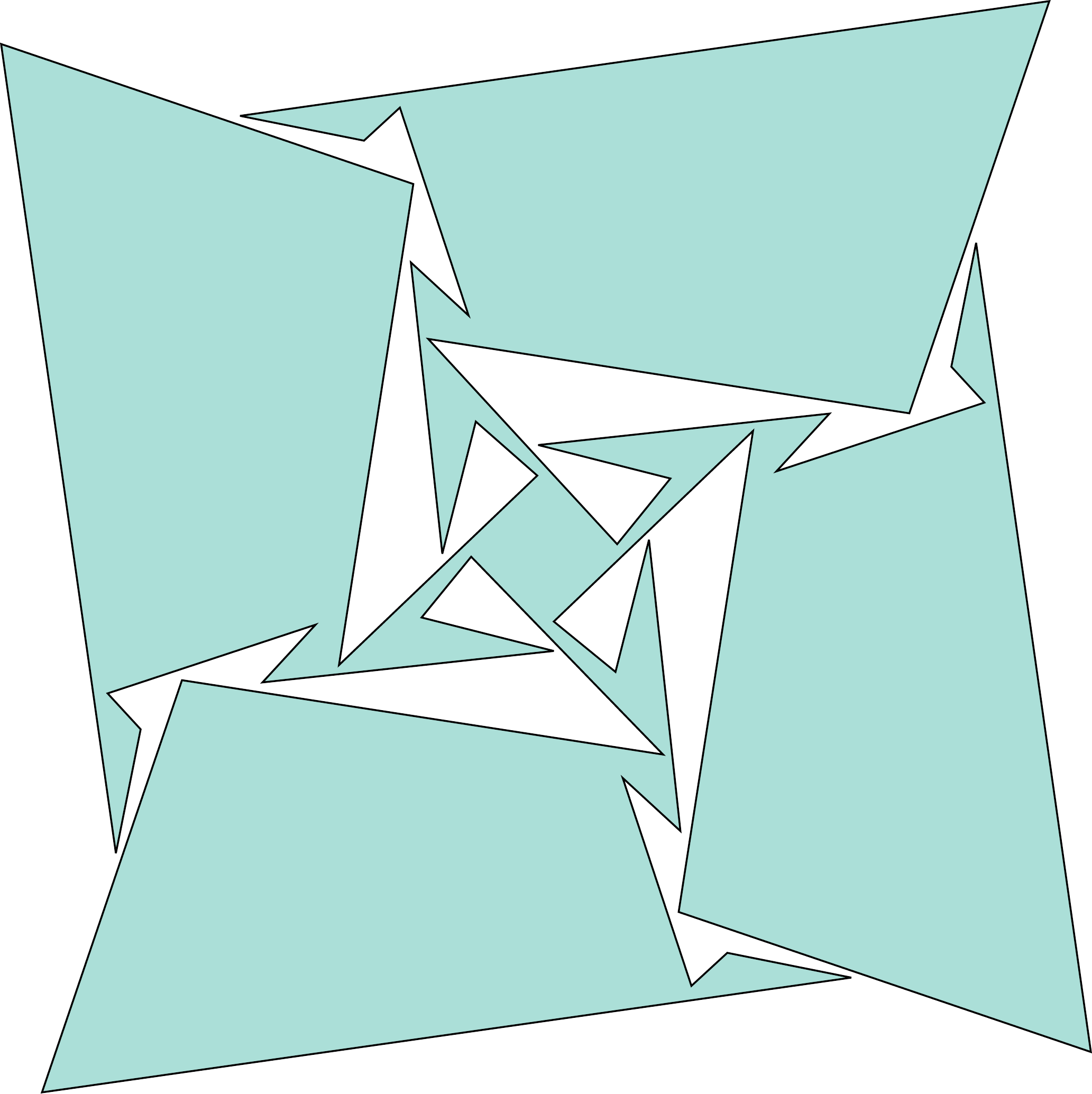}
\end{tabular}
\vspace{-0.5em}\caption{(a) \hp$(ab, v)$ illustrated (b) Polygon
admitting no \hp$\mbox{s}$.} \figlab{hops}
\end{figure}

The limited transformation capabilities of the swap and hop
operations motivate our introduction of two new operations,
\emph{stretch} and \emph{twang}. The former operation relaxes the two hop
conditions and allows the creation of a polygonal wrap.
The latter operation restores the polygonal wrap to a polygon. We show
that together they are capable of transforming any polygon into a
canonical form
(Sections~\secref{single.pocket.reduction}-\secref{canonical}), and
from there to any other polygon
(Sections~\secref{Reverse}-\secref{Connectivity}).

\subsection{Stretches and Twangs}
\seclab{Stretches.Twangs}

Unlike the \hp$(e, v)$ operation, which requires $v$ to fully see
the edge $e$ into which it is hopping, the \st$(e, v)$ operation
only requires that $v$ see a point $x$ in the interior\footnote{
   By ``interior'' we mean ``relative interior,''
   i.e., not an endpoint.} of $e$.
The stretch is accomplished in two stages: (i)~temporarily
introduce two new ``pseudovertices'' on $e$ in a small neighborhood
of $x$ (this is what we call \st$_0$ below), and (ii)~remove the
pseudovertices immediately using twangs.

\paragraph{\st$_0$.}
Let $v$ see a point $x$ in the interior of an edge $e$ of $P$. By
\emph{see} we mean ``clear visibility'', i.e., the segment $vx$
shares no points with $\partial P$ other than $v$ and $x$ (see
Fig.~\ref{fig:stretch.twang}a). Note that every vertex $v$ of $P$
sees such an $x$ (in fact, infinitely many $x$) on some $e$.
Let $x^-$ and $x^+$ be two points to either side of $x$ on $e$,
both in the interior of $e$,
such that $v$ can clearly see both $x^-$ and $x^+$.
Two such points always exist in a
neighborhood of $x$. We call these points \emph{pseudovertices}. Let
$e=ab$, with $x^-$ closer to the endpoint $a$ of $e$. Then
\st$_0(e,v)$ alters the polygon to replace $e$ with $(a, x^-, v,
x^+, b)$, effectively ``stretching'' $e$ out to reach $v$ by
inserting a narrow triangle $\triangle x^- v x^+$ that
sits on $e$ (see Fig.~\figref{stretch.twang}b).

\begin{figure}[htbp]
\centering
\includegraphics[width=0.9\linewidth]{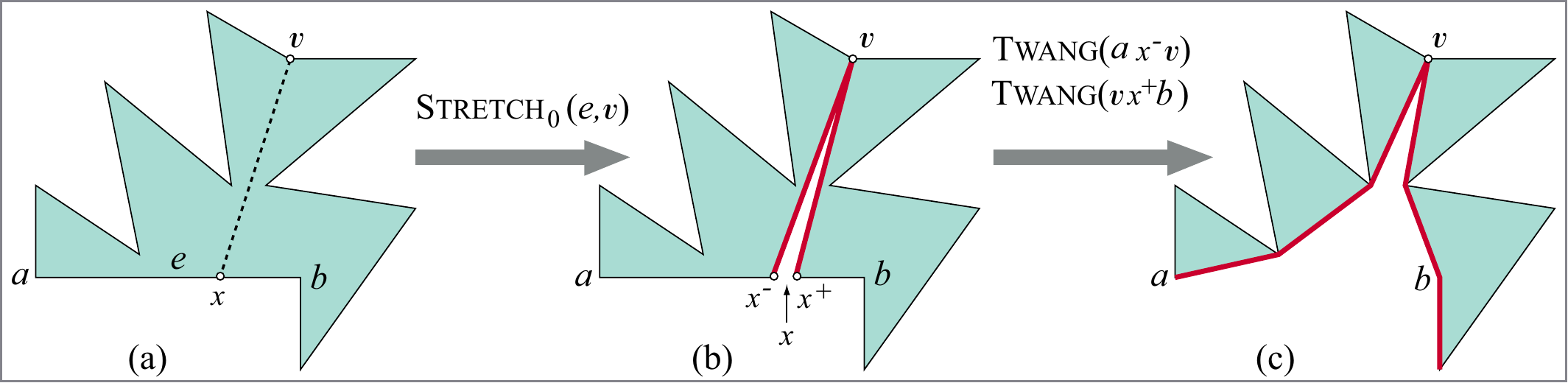}
\caption{\st$(e, v)$ illustrated (a) $v$ sees $x \in e$ (b)
\st$_0(e,v)$ (c) \st$(e,v)$. } \figlab{stretch.twang}
\end{figure}
%
To complete the definition of \st$(e,v)$, which
removes the pseudovertices 
$x^+$ and $x^-$, we 
first define the twang operation.

\vspace{-0.5em}
\paragraph{\tw.}
Informally, if one views the polygon boundary as an elastic band, a
twang operation detaches the boundary from a vertex $v$ and snaps it
to $v$'s convex side. 

\begin{definition}
\emph{The operation \tw$(abc)$ is defined for any three consecutive
vertices $abc \in \sigma$ such that
\begin{enumerate}
\squeezelist
\item $\{ a,b,c \}$ are not collinear.  
\item $b$ is either a pseudovertex, or a vertex in double
contact. If $b$ is a vertex in double contact,  
then $\triangle abc$
does not contain a \emph{nested} double contact at $b$. 
By
this we mean the following: Slightly perturb the vertices of $P$ to
separate each double-contact into two or more points, so that $P$ becomes
simple. Then $\triangle abc$ does not contain any other occurrence of $b$ in
$\sigma$. (E.g., in Fig.~\figref{twangs}a, $\triangle a'bc'$ contains a second occurrence
of $b$.)
\end{enumerate}
}

\emph{
Under these conditions, the operation \tw$(abc)$ replaces the
sequence $abc$ in $\Pw$ by $\sp(abc)$, where $\sp(abc)$ indicates
the shortest path from $a$ to $c$ that stays inside $\triangle abc$
and does not cross $\partial \Pw$. We call $b$ the 
\emph{twang vertex}. Whenever $a$ and $c$ are irrelevant to the discussion, we
denote the twang operation by \tw$(b)$.}  \label{def:twang}
\end{definition}


%
%

\begin{figure}[htbp]
\centering
\includegraphics[width=\linewidth]{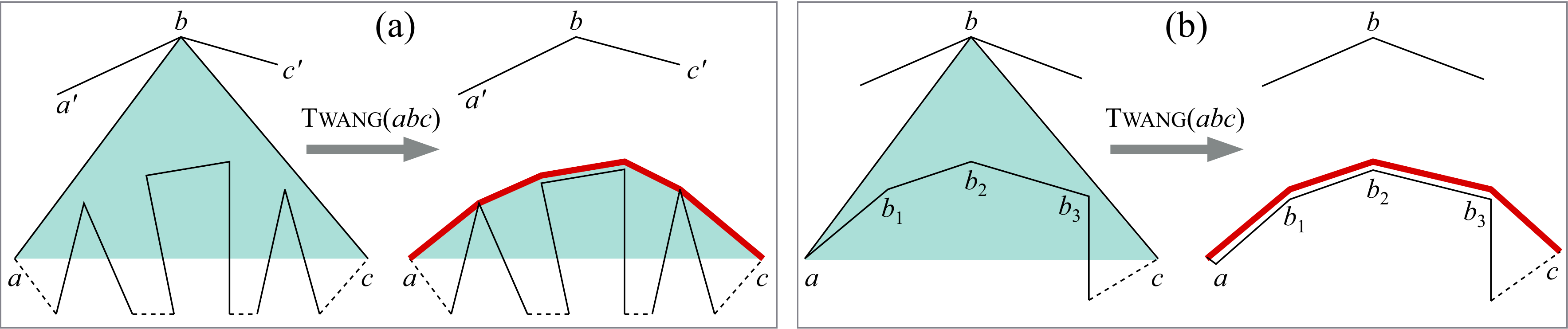}
\caption{\tw($abc$) illustrated (a) \tw$(abc)$ replaces $abc$ by
$\sp(abc)$ (b) \tw$(abc)$ creates the hairpin vertex $a$ and three
doubled edges $ab_1$, $b_1b_2$ and $b_2b_3$.} \figlab{twangs}
\end{figure}

Informally, \tw$(abc)$ ``snaps'' the boundary to wrap around the hull
of the points in $\triangle abc$, excluding $b$ (see
Fig.~\figref{twangs}a). A twang operation can be viewed as taking
a step toward simplicity by removing either a pseudovertex or a point
of double contact.  We should note that $\sp(abc)$ includes every
vertex along this path, even collinear vertices. If there are no
points inside $\triangle abc$, then $\sp(abc)=ac$, and \tw$(abc)$ can
be viewed as the reverse of \hp$(ac,b)$.
If $a{=}c$ (i.e., $ab$ and $bc$ overlap in $\Pw$), we call $b$ a \emph{hairpin}
vertex of $\Pw$; in this case, \tw$(aba)$ replaces $aba$ in $\Pw$ by
$a$. Hairpin vertices and ``doubled edges'' 
arise naturally from twangs.
In Fig.~\figref{twangs}b for instance, \tw$(abc)$ produces a
hairpin vertex at $a$ and doubled edges $ab_1$, $b_1b_2$, $b_2b_3$.
So we must countenance such degeneracies.
In general, there are points interior to the triangle, and the twang
creates new points of double contact.  Below, we will apply twangs
repeatedly to remove all double contacts.

\paragraph{\st.}
We can now complete the definition of \st$(e,v)$, with $e=ab$.  
First execute
\st$_0(e,v)$, which
picks the two pseudovertices $x^+$ and $x^-$.
Then execute \tw$(ax^-v)$ and
\tw$(vx^+b)$, which detach the boundary from
$x^+$ and $x^-$ and return to a polygonal wrap 
of 
$S$ (see Fig.~\figref{stretch.twang}c). We refer to $e$
($v$) as the \emph{stretch edge} (\emph{vertex}).

\vspace{-0.2em}
\subsection{Twang Cascades}
\seclab{Twang.Cascades}
A twang in general removes one double contact and creates perhaps
several others.  A \emph{\twc} applied on a polygonal wrap
$\Pw$ removes all points of double contact from $\Pw$:

\vspace{-1em}
\begin{center}
\vspace{1mm} \fbox{
\begin{minipage}[h]{0.98\linewidth}
\centerline{\twc($\Pw$)}
\vspace{1mm}{\hrule width\linewidth}\vspace{2mm} 
\small{\begin{tabbing}
...\=..........\=...........................................................\kill
Loop for as long as $\Pw$ has a point of double contact $b$:\\
\\
\> 1. Find a vertex sequence $abc$ in $\Pw$ that satisfies the twang conditions (cf. Def.~\ref{def:twang}).\\
\> 2. \tw$(abc)$.
\end{tabbing}}
\end{minipage}
}
\vspace{1mm}
\end{center}

%

Note that for any point of double-contact $b$, there always exists a vertex
sequence $abc$ that satisfies the twang conditions and therefore the twang
cascade loop never gets stuck.
That a twang cascade eventually terminates is not immediate.
%
The lemma below, whose proof we omit, shows that \tw$(abc)$ shortens
the perimeter of the polygonal wrap 
(because it replaces $abc$ by $\sp(abc)$)
by at least a constant depending
on the geometry of the point set.  
Therefore, any twang cascade must terminate in a finite number of steps.

\begin{lemma}
A single twang \tw($abc$) decreases the perimeter of the polygonal wrap by at least
$2d_{\min} (1 - \sin(\alpha_{\max}/2))$, where $d_{\min}$ is the
smallest pairwise point distance and $\alpha_{\max}$ is the maximum convex
angle formed by any triple of non-collinear points.
\lemlab{delta}
\end{lemma}

Supplementing this geometric bound,
Corollary~\corref{odometer} in Appendix~3 establishes 
a combinatorial upper bound of $O(n^n)$ on the number of twangs
in any twang cascade.
An impediment to establishing a better bound is that a point can
twang more than once in a cascade.
Indeed we present in Appendix~2 an example in which
$\Omega(n)$ points each
twang $\Omega(n)$ times in one cascade, providing an
$\Omega(n^2)$ lower bound.

\subsubsection{Forward Move}
We define a \emph{forward move} on a polygonization $P$ of a set $S$
as a stretch (with the additional requirement that the
pseudovertices on the stretch edge lie on the reflex side of the
stretch vertex), followed by a twang and then a twang cascade, as
described below:

\begin{center}
\vspace{1mm} \fbox{
\begin{minipage}[h]{0.95\linewidth}
\centerline{\fm($P, e,v$)}
\vspace{1mm}{\hrule width\linewidth}\vspace{2mm} 
\small{\begin{tabbing}
..\=.....\=...........................................................\kill
\> Preconditions: (i) $P$ is a simple polygon, (ii) $e$ and $v$
satisfy the conditions of \st$(e,v)$, and \\ 
\> (iii) $v$ is a noncollinear
vertex such that pseudovertices $x^+$ and $x^-$ on $e$ lie on the reflex
side of $v$.\\ 
\> \{Let $u, v, w$ be the vertex sequence containing $v$ in $P$
(necessarily unique, since $P$ is simple).\}\\ 
\\
\> 1. \> $\Pw \leftarrow$ \st$(e, v)$. \\
\> 2. \> $\Pw \leftarrow$ \tw$(uvw)$. \\
\> 3. \> $P' \leftarrow$ \twc$(\Pw)$.
\end{tabbing}}
\end{minipage}
}
\vspace{1mm}
\end{center}


A \fm\ takes one polygonization $P$ to another $P'$ (see
Fig.~\figref{twang.cascade}), as follows from
Lemma~\lemref{delta}.
Next we discuss two important phenomena that can occur
during a forward move.

\vspace{-0.5em}
\paragraph{Stretch Vertex Placement.} We note that the initial
stretch that starts a move might be ``undone''
by cycling of the cascade. This phenomenon is illustrated in
Fig.~\figref{twang.cascade}, where the initial \st$(ab,v)$ inserts
$v$ between $a$ and $b$ in the polygonal wrap
(Fig.~\figref{twang.cascade}b), but $v$ ends up between $c$ and $b$
in the final polygonization (Fig.~\figref{twang.cascade}f). Thus any
attempt to specifically place $v$ in the polygonization sequence
between two particular vertices might be canceled by the subsequent
cascade. This phenomenon presents a challenge to reducing a polygon
to canonical form (discussed in Section~\secref{canonical}).

\vspace{-1.5em}
\begin{figure}[htbp]
\centering
\includegraphics[width=\linewidth]{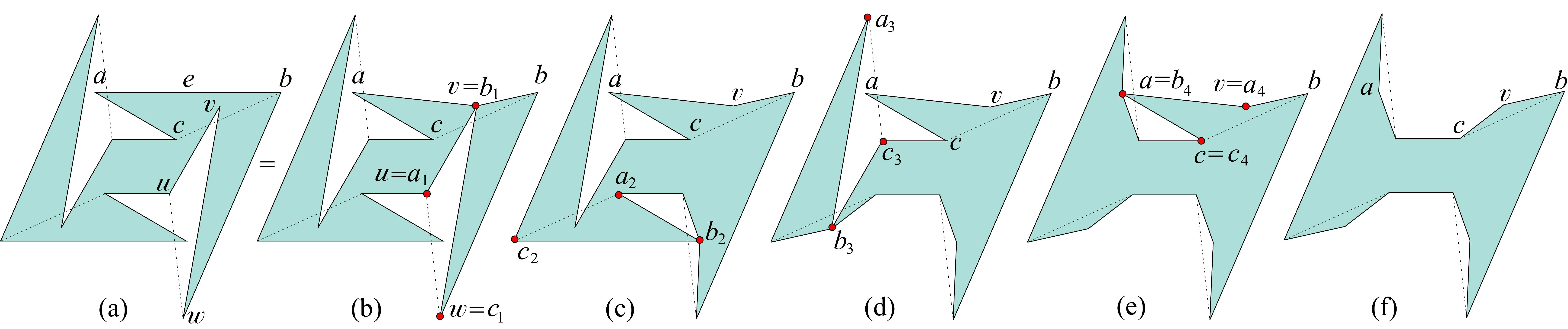}
\vspace{-1.7em} \caption{Forward move illustrated. (a) Initial
polygon $P$ (b) After \st$(ab, v)$ (c) After \tw$(a_1b_1c_1)$ (d)
After \tw$(a_2b_2c_2)$ (e) After \tw$(a_3b_3c_3)$ (f) After
\tw$(a_4b_4c_4)$.\vspace{-0.7em}} \figlab{twang.cascade}
\end{figure}


\vspace{-1.5em}
\section{Single Pocket Reduction Algorithm}
\seclab{single.pocket.reduction} Now that the basic properties of
the moves are established, we aim to show that our moves suffice to
connect any two polygonizations of a point set $S$. The plan is to
reduce an arbitrary polygonization to the canonical polygonization.
En~route to explaining this reduction algorithm, we show how to
remove any particular pocket by redistributing its vertices to other
pockets. This method will be applied repeatedly in
Section~\secref{multiple.pocket.reduction} to move all pockets to
one particular pocket.

In this section we assume that $P$ has two or more
pockets. 
We use $\hull(P)$ to refer to the closed region defined by
the convex hull of $P$. For a fixed hull edge $e$ that is the lid of a
pocket $A$, the goal is to reduce $A$ to $e$ by redistributing the
vertices of $A$ among the other pockets, using forward moves
only. 
This is accomplished by the
{\sc Single Pocket Reduction}~algorithm (described below), which
repeatedly picks a hull vertex $v$ of $A$ and attaches $v$ to a pocket
other than $A$; see Fig.~\figref{single.pocket.reduction} for an
example run. 
Call a vertex $v$ of $P$ a \emph{true} corner if the two polygon edges
incident to $v$ are
non-collinear. 

\begin{center}
\vspace{1mm} \fbox{
\begin{minipage}[h]{0.95\linewidth}
\centerline{{\sc Single Pocket Reduction}($P, e$) Algorithm}
\vspace{1mm}{\hrule width\linewidth}\vspace{2mm} 
\small{\begin{tabbing}
.....\=..........\=...........................................................\kill
Loop for as long as the pocket $A$ of $P$ with lid $e$ contains three or more vertices:\\
\> 1. Pick an edge-vertex pair $(e, v)$ such that \\
\> \> $e$ is an edge of $P$ on $\partial B$ for some pocket $B \neq A$\\
\> \> $v\in A$ is a non-lid true corner vertex on $\hull(A)$ that sees $e$\\
\> 2. $P \leftarrow$ \fm$(P, e,v)$.
\end{tabbing}}
\end{minipage}
}
\vspace{1mm}
\end{center}

We now establish that the {\sc Single Pocket
Reduction} algorithm 
terminates in a finite number of iterations. First we 
prove a
more general lemma
showing that a twang operation can potentially reduce, but never
expand, the hull of a pocket.

\begin{lemma}[Hull Nesting under Twangs]
Let $A$ be a pocket of a polygonal wrap $\Pw$ and let vertex $b
\not\in \hull(\Pw)$ satisfy the twang conditions.
Let $A'$ be the pocket with the
same lid as $A$ after
\tw$(b)$. Then $A' \subseteq \hull(A)$.
\lemlab{twang.hull}
\end{lemma}

\begin{pf}
Let $abc$ be the vertex sequence involved in the twang operation.
Then \tw$(abc)$ replaces the path $abc$ by $\sp(abc)$.
If $abc$ does not belong to $\partial A$,
then \tw$(abc)$ does not affect $A$ and therefore $A' \equiv A$. So assume that
$abc$ belongs to $\partial A$. This implies that $b$ is a vertex of $A$.
Note that $b$ is a non-lid vertex, since
$b \not\in \hull(\Pw)$. Then $\triangle abc \subset \hull(A)$, and
the claim follows from the fact that $\sp(abc) \subset \triangle abc$.
\end{pf}

\vspace{-1.0em}
\begin{figure}[htbp]
\centering
\includegraphics[width=\linewidth]{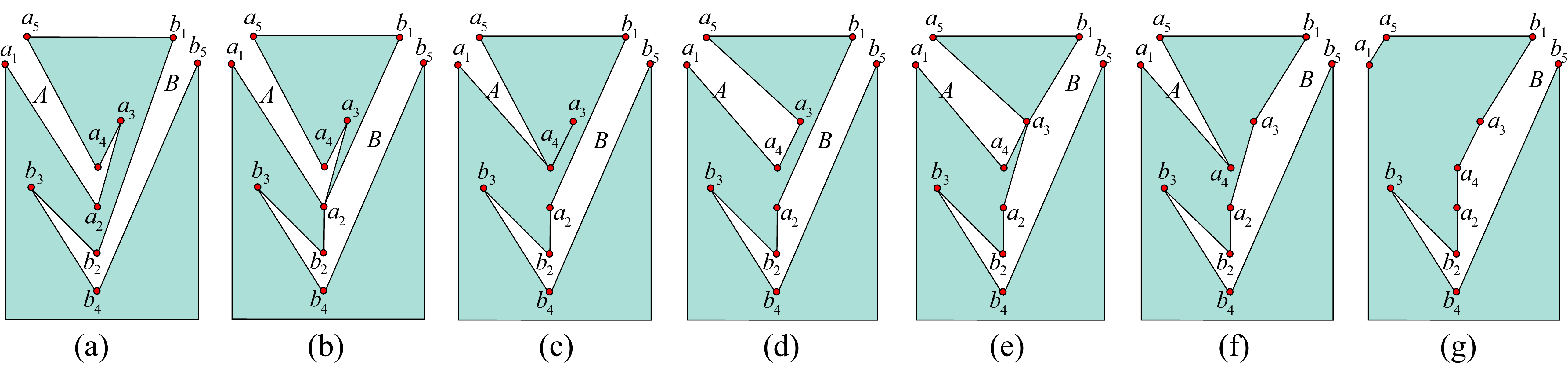}
\vspace{-2em}\caption{\small{{\sc Single Pocket Reduction}($P,
a_1a_5$) 
illustrated: (a) Initial $P$; (b) After \st$(b_1b_2, a_2)$;
(c) After \tw$(a_1a_2a_3)$; (d) After \tw$(a_3a_4a_5)$; (e) After
\st$(a_2b_1, a_3)$; (f) After \tw$(a_4a_3a_5)$; (g) After
\st$(a_2a_3, a_4)$+\tw$(a_1a_4a_5)$.}}
\figlab{single.pocket.reduction}
\end{figure}


\vspace{-0.8em}
\begin{lemma}
The {\sc Single Pocket Reduction} algorithm terminates in $O(n)$ forward moves.
\lemlab{single.pocket.reduction}
\end{lemma}
\begin{pf}
Let $S$ denote the set of vertices of $P$ in $\hull(A)$. Thus $|S| =
O(n)$. We show that $|S|$ decreases by at least $1$ in each loop
iteration, thus establishing the claim of the lemma.

First observe that the existence of an edge-vertex pair $(e, v)$
  selected in Step 1 is guaranteed by the fact that $P$ has two or
  more pockets.
%
%
Step 2 of the {\sc Single Pocket Reduction} algorithm, which performs
a forward move to a different polygonization,
attempts to reduce $A$ by vertex $v$, thus decrementing $|S|$. We now show
that this step is successful in that it
does not reattach $v$ back to $A$. Furthermore, we show that
$S$ acquires no new vertices during this step. These together
show that $|S|$ decreases by at least 1 in each loop iteration.

The first step of the forward move, \st$(e, v)$, does not affect $S$.
The second step, \tw$(avb)$, replaces
the path $avb$ by $\sp(avb)$, thus eliminating $v$ from
$A$. Since $v$ is a true corner vertex of $A$, $\hull(A)$
does not contain $v$ at the end of this step. Let $A'$ be the pocket of $P$
with the same lid as $A$ at the end of 
\twc$(P)$. Since a hull vertex never twangs,
Lemma~\lemref{twang.hull} implies that
$\hull(A')$ is a subset of $\hull(A)$ and therefore $|S|$ does
not increase during the twang cascade. Furthermore, since
$\hull(A)$ does not contain $v$ after the first twang operation,
$v$ must lie outside of $\hull(A')$ at the end of the twang
cascade. 
\end{pf}


\vspace{-0.4em}
\section{Multiple Pocket Reduction Algorithm}
\seclab{multiple.pocket.reduction} For a given hull edge $e$, the
goal is to transform $P$ to a polygon with a single pocket with lid
$e$, using forward moves only. If $e$ is an edge of the polygon, for
the purpose of the algorithm discussed here we treat $e$ as a
(degenerate) target pocket $T$. We assume that, in addition to
$T$,~$P$ has one or more other pockets, otherwise there is nothing to do.
Then we can use the {\sc Single Pocket Reduction} algorithm to
eliminate all pockets of $P$ but $T$, as described in the {\sc
Pocket Reduction} algorithm below.

\vspace{-0.2em}
\begin{center}
\vspace{1mm} \fbox{
\begin{minipage}[h]{0.95\linewidth}
\centerline{{\sc Pocket Reduction} ($P, e$) Algorithm}
\vspace{1mm}{\hrule width\linewidth}\vspace{2mm} 
\small{\begin{tabbing}
.....\=.....\=...........................................................\kill
If $e$ is an edge of $P$, set $T \leftarrow e$, otherwise set $T \leftarrow$ the pocket with lid $e$ \\ \>(in either case, we treat $T$ as a pocket). \\
For each pocket lid $e' \neq e$ \\
\> Call {\sc Single Pocket Reduction}($P, e'$)
\end{tabbing}}
\end{minipage}
}
\vspace{1mm}
\end{center}
%


\vspace{-0.2em} Observe that the {\sc Pocket Reduction} algorithm
terminates in $O(n^2)$ forward moves: there are $O(n)$ pockets 
each of which gets reduced to its lid edge in $O(n)$ forward moves
(cf. Lemma~\lemref{single.pocket.reduction}).

Fig.~\figref{manytoone.example} illustrates the {\sc Pocket
Reduction} algorithm on a 17-vertex polygon with three pockets $A$,
$B$ and $C$, 
each of which has 3 non-lid vertices,
and target pocket $T$ with lid edge $e = t_1t_2$. 
The algorithm first
calls {\sc Single Pocket Reduction}($P, a_1a_5$), which transfers to
$B$ all non-lid vertices of $A$, so $B$ ends up with 6 non-lid
vertices (this reduction is illustrated in detail in
Fig.~\figref{single.pocket.reduction}). Similarly, {\sc Single
Pocket Reduction}($P, b_1b_5$) transfers to $C$ all non-lid vertices
of $B$, so $C$ ends up with 9 non-lid vertices, and finally {\sc
Single Pocket Reduction}($P, c_1c_5$) transfers all these vertices
to $T$.

\begin{figure}[htbp]
\centering
\includegraphics[width=\linewidth]{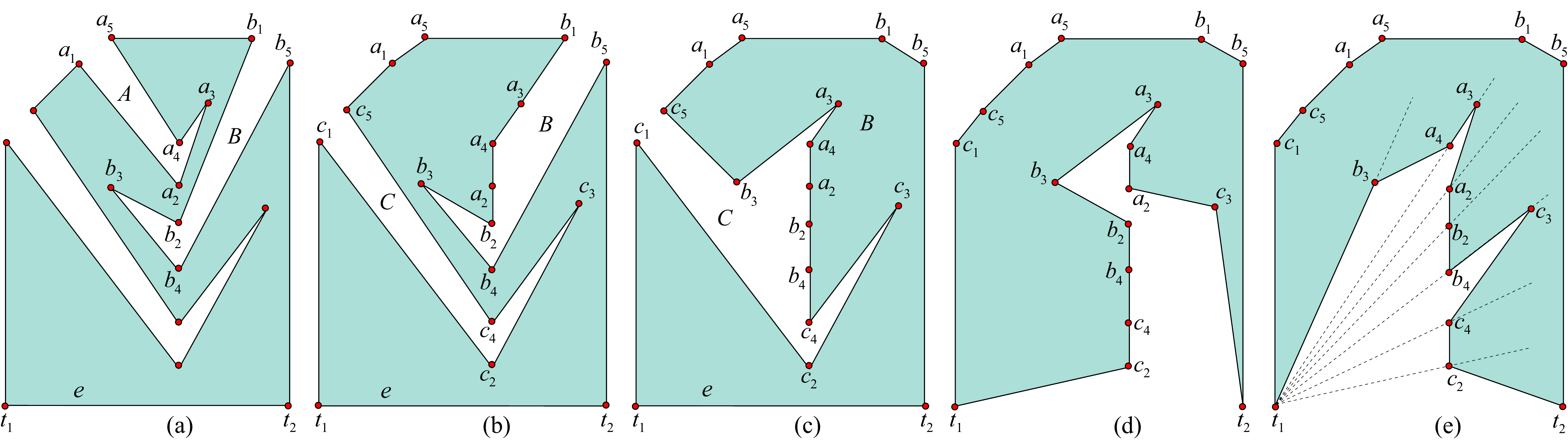}
\vspace{-2em}\caption{(a-e) \small{{\sc Pocket Reduction}($P,
t_1t_2$): (a) Initial $P$; (b) After {\sc Single Pocket
Reduction}($P, a_1a_5$); (c) After {\sc Single Pocket Reduction}($P,
b_1b_5$); (d) After {\sc Single Pocket Reduction}($P, c_1c_5$);
(e) After {\sc Canonical Polygonization}($P, t_1t_2$).}}
\figlab{manytoone.example}
\end{figure}

This example shows that the $O(n^2)$ bound on the number of forward
moves is tight: an $n$-vertex polygon with a structure similar to
the one in Fig.~\figref{manytoone.example}a has $O(n)$
 pockets.
The number of forward moves performed
by the {\sc Pocket Reduction} algorithm is therefore $ 3 + 6 + 9 +
\ldots \frac{3n}{5} = \Theta(n^2) $, so we have the following lemma:
\begin{lemma}
The {\sc Pocket Reduction} algorithm employs $\Theta(n^2)$ forward
moves. \lemlab{multiple.pocket.bound}
\end{lemma}

\vspace{-1.2em}
\section{Single Pocket to Canonical Polygonization}
\seclab{canonical} Let $P(e)$ denote an arbitrary one-pocket
polygonization of $S$ with pocket lid $e=ab$.  Here we give an
algorithm to transform $P(e)$ into the canonical polygonization
$P_c(e)$. This, along with the algorithms discussed in
Sections~\secref{single.pocket.reduction}
and~\secref{multiple.pocket.reduction}, gives us a method to
transform any polygonization of $S$ into the canonical form
$P_c(e)$. 
Our canonical polygonization algorithm incrementally arranges pocket
vertices in canonical order (cf. Section~\secref{Pockets}) along the
pocket boundary
by applying a series of 
forward moves to $P(e)$.

\begin{center}
\vspace{1mm} \fbox{
\begin{minipage}[h]{0.95\linewidth}
\centerline{\sc Canonical Polygonization($P, e$) Algorithm}
\vspace{1mm}{\hrule width\linewidth}\vspace{2mm} 
     {\small
\begin{tabbing}
.....\=.....\=..........................................................\kill
Let $e = ab$.
Let $a = v_0, v_1, v_2, \ldots, v_k, v_{k+1} = b$ be the canonical order of
the vertices of pocket $P(e)$.\\
For each $i = 1, 2, \ldots, k$\\
\> 1. Set $\ell_i \leftarrow$ line passing through $a$ and $v_i$ \\
\> 2. Set $e_{i-1} \leftarrow$ pocket edge $v_{i-1}v_j$, with $j > i-1$ \\
\> 3. If $e_{i-1}$ is not identical to $v_{i-1}v_i$, apply \fm($e_{i-1}, v_i$).
\end{tabbing}}
\end{minipage}
}
\vspace{1mm}
\end{center}

We now show that the one-pocket polygonization resulting after the
$i$-th iteration of the loop above has the points $v_0, \ldots, v_i$
in canonical order along the pocket boundary. This, in turn, is
established by showing that the \fm\ in the $i$-th iteration
involves only points in the set $\{v_i, v_{i+1},\ldots, v_k\}$.
These observations are formalized in the following lemma, whose
proof appears in Appendix~1:

\begin{lemma}
The $i$-th iteration of the {\sc Canonical Polygonization} loop
produces a polygonization of $S$ with one pocket with lid $e$ and
with vertices $v_0,\ldots, v_i$ consecutive along the pocket
boundary. \lemlab{canonical.first}
\end{lemma}

\begin{lemma}
The {\sc Canonical Polygonization} algorithm constructs $P_c(e)$ in
$O(n)$ forward moves. \lemlab{canonical.final}
\end{lemma}

\vspace{-1.5em}
\section{Reverse Moves}
\seclab{Reverse}
Connectivity of the space of polygonizations will follow by reducing
two given polygonizations $P_1$ and $P_2$ to a common canonical form
$P_c$, and then reversing the moves from $P_c$ to $P_2$.
Although we could just define a reverse move as a
time-reversal of a forward move, it must be admitted that such
reverse moves are less natural than their forward counterparts.  So
we concentrate on establishing that reverse moves can be achieved by
a sequence of atomic stretches and twangs.

\vspace{-0.5em}
\paragraph{Reverse Stretch.} The reverse of \st$(e,v)$ may be achieved by a
sequence of one or more twangs, as illustrated in
Fig.~\figref{unstretchtwang}a. This result follows from the fact
that the ``funnel'' created by the stretch is empty, and so the
twangs reversing the stretch do not cascade.

\begin{figure}[htbp]
\centering
\includegraphics[width=\linewidth]{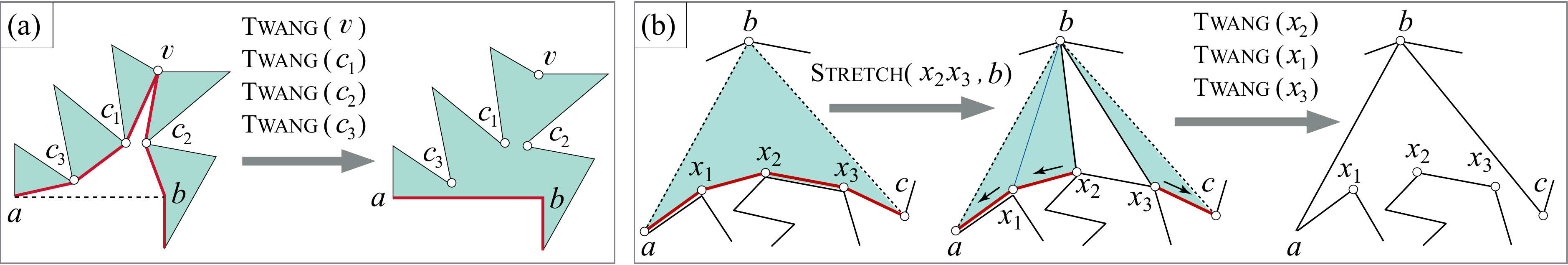}
\caption{Reverse atomic moves: (a) \st$(ab, v)$ is reversed by
\tw$(v)$, \tw$(c_1)$, \tw$(c_2)$, \tw$(c_3)$. (b) \tw$(b)$ is
reversed by \st$(x_2x_3,b)$, \tw$(x_2)$, \tw$(x_1)$ and \tw$(x_3)$.}
\figlab{unstretchtwang}
\end{figure}


\vspace{-1.5em}
\paragraph{Reverse Twang.} 
An ``untwang'' can be accomplished by one stretch followed by a series
of twangs. Fig.~\figref{unstretchtwang}b illustrates how \tw$(abc)$
may be reversed by one \st$(e,b)$, for any edge $e$ of $\sp(abc)$,
followed by zero or more twangs. Observe that the initial stretch
in the reverse twang operation is not restricted to the reflex side of
the stretch vertex, as it is in a \fm. If $b$ is
a hairpin vertex (i.e., $a$ and $c$ coincide), we view $ac$ as an edge
of length zero and the reverse of \tw$(b)$ is simply \st$(e,b)$.


\vspace{-0.5em}
\paragraph{Consequence.}
We have shown that the total effect of any forward move, consisting
of one stretch and a twang cascade, can be reversed by a sequence of
stretches and twangs. We call this sequence a \emph{reverse
move}. One way to view the consequence of the above two results can
be expressed via regular expressions. Let the symbols $s$ and $t$
represent a \st\ and \tw\ respectively. Then a forward move can be
represented by the expression $s t^+$: a stretch followed by
one or more twangs. 
A reverse stretch, $s^{-1}$ can be achieved by one or more twangs:
$t^+$. And
a reverse twang $t^{-1}$ can be achieved by $s t^*$. Thus the
reverse of the forward move $s t^+$ is $(t^{-1})^+ s^{-1} = (s
t^*)^+ t^{+} \;,$ 
 a sequence of stretches and twangs, at least one of each.

\vspace{-0.5em}
\section{Connectivity and Diameter of Polygonization Space}
\seclab{Connectivity}
We begin with a summary the algorithm which, given two polygonizations $P_1$
and $P_2$ of a fixed point set, transforms $P_1$ into $P_2$ using
stretches and twangs only.

\vspace{-0.2em}
\begin{center}
\vspace{1mm} \fbox{ \small{
\begin{minipage}[h]{0.95\linewidth}
\centerline{{\sc Polygon Transformation}($P_1, P_2$) Algorithm}
\vspace{1mm}{\hrule width\linewidth}\vspace{2mm} 
     {\small
\begin{tabbing}
.....\=.....\=..........................................................\kill
1. Select an arbitrary edge $e$ of $\hull(P_1)$. \\
2. $P_1 \leftarrow$ {\sc Pocket Reduction}($P_1, e$); $M_1 \leftarrow$ atomic moves of [$P_2 \leftarrow$ {\sc Pocket Reduction}($P_2, e$)].\\
3. $P_c \leftarrow$ {\sc Canonical Polygonization}($P_1, e$); $M_2 \leftarrow$ atomic moves of [{\sc Canonical Polygonization}($P_2, e$).]\\
4. Reverse the order of the moves in $M_1 \oplus M_2$ ($\oplus$ represents concatenation). \\
5. For each stretch $s$ (twang $t$) in $M_1 \oplus M_2$ in order,
execute reverse stretch $s^{-1}$(reverse twang $t^{-1})$ on $P_c$.
\end{tabbing}}
\end{minipage}
}}
\vspace{1mm}
\end{center}
%

\vspace{-0.2em} \noindent This algorithm, along with
Lemmas~\lemref{multiple.pocket.bound} and \lemref{canonical.final},
establishes our main theorem:

\vspace{-0.2em}
\begin{theorem}
The space of polygonizations of a fixed set of $n$ points is connected via a
sequence of forward and reverse moves.
The diameter of the polygonization space is $O(n^2)$ moves.
\end{theorem}

\vspace{-1.2em}
\paragraph{Computational Complexity.} With appropriate preprocessing, each twang
operation can be carried out in $O(n)$ time (since $\sp()$ might hit $O(n)$ vertices).
So the running time of a single forward/reverse move is $T \cdot O(n)$, where $T$ is an
upper bound on the number of twangs in a move. 

\section{Random Polygons}
\seclab{Random.Polygons}
Let $G$ be the graph whose nodes are polygonizations and whose
arcs are the 
moves defined in this paper.
We know that $G$ can have an exponential number $N$ of vertices.
We have established that it is connected,
and that it has diameter $O(n^2)$.
One way to generate ``random polygons,'' as mentioned in
Section~\secref{Introduction}, is to start with some polygonization
$P$ of a random set of $n$ points $S$,
and repeatedly select moves randomly.
An immediate question here is: How
many moves are needed to achieve adequate mixing, i.e.,
ergodicity? 
If we view the random moves as a random walk in $G$,
and ask for the 
expected time for a random walk to visit all $N$
vertices (the \emph{expected cover time})
of a connected graph $G$,
the answer is known: $\Theta(N^3)$~\cite{f-tupct-95}.
Unfortunately, 
the maximum number of polygonizations of a set of $n$ points
has been shown to be 
$\Omega(4.6^n)$~\cite{gnt-lbncf-00},
and it seems likely that the expected number of polygonizations of a
random set of points is also exponential
(although
we have not found this established in the literature).
Thus, exponentially many moves
are needed to cover the polygonization graph in the worst case, and perhaps
in the expected case as well.
Although disappointing, this is inevitable given the size of $G$,
and mitigated somewhat by the relatively small diameter of $G$.


We have implemented a version of random polygon generation. After
creating an initial polygonization, we move from polygonization to
polygonization via a sequence of forward moves, where additional
stretches are permitted in the cascade to simulate reverse moves.
Trials on random polygons suggest that the average length of a
cascade for polygons of up to $n{=}100$ vertices is about $1.3$, with
$7$ the maximum cascade length observed in
trials of thousands of forward moves. 
Cascade length seems to be independent of $n$.
Thus, even though 
we only have loose bounds on the length of a twang cascade,
for random point sets the mean length appears to be a constant less than $2$.

\vspace{-0.8em}
\section{Open Problems}
\vspace{-0.2em} 
\seclab{Open.Problems} Our work leaves many interesting problems
open. 
The main unresolved question is establishing a tighter combinatorial
bound on the number of twangs $T$ in a twang cascade and thereby
resolving the computational complexity of the polygon transformation
algorithm.  We have shown (in Appendices)
that $T$ is $\Omega(n^2)$ and $O(n^n)$,
leaving a large gap to be closed.
Another related question asks to improve the efficiency of
the polygon transformation algorithm in terms of forward moves (the
lower bound in Lemma~\lemref{multiple.pocket.bound} is for our
particular algorithm, not all algorithms).

In Section~\secref{Connectivity} we established connectivity with
forward moves and their reverse, and although both moves are
composed of atomic stretches and twangs, the forward moves seem more
naturally determined. 
This suggests the question of
whether forward moves
suffice to ensure
connectivity.
%
It remains to be seen if the polygonization moves explored in this
paper will be effective tools for generating random polygons.
%
One possibility is to start from a doubled random noncrossing spanning
tree, which is a polygonal wrap.
%
%
Finally, we are extending our work to 3D 
polyhedralizations of a fixed 3D point set.



\normalsize

\newpage
\section*{Appendix~1: {\sc Canonical Polygonization}}
\begin{lemma}
The $i$-th iteration of the {\sc Canonical Polygonization} loop
produces a polygonization of $S$ with one pocket with lid $e$ and
with vertices $v_0,\ldots, v_i$ consecutive along the pocket
boundary.
\end{lemma}
\begin{pf}
The proof is by induction.
The base case corresponds to $i=1$ and is trivially true for the
case when $e_0 = v_0v_1$. Otherwise, $v_1$ sees $e_0$ (since no edge
can block visibility from $v_0$ to $v_1$) and therefore \st$(e_0,
v_1)$ 
is possible. See
Fig.~\figref{canonical.proof.firstfig}a. Furthermore, $v_1$ may not
twang a second time during the twang cascade 
of the forward move. This is because a second \tw$(v_1)$ may only be
triggered by the twang of a hull vertex, which can never occur (hull
vertices never twang). This implies that the \fm\ in Step 3 of the
first iteration creates a one-pocket polygonization in which $v_0$
and $v_1$ are consecutive along the pocket boundary (see
Fig.~\figref{canonical.proof.firstfig}a,b). This completes the base
case.

\begin{figure}[htbp]
\centering
\begin{tabular}{c@{\hspace{0.01\linewidth}}c@{\hspace{0.01\linewidth}}c}
\includegraphics[width=0.32\linewidth]{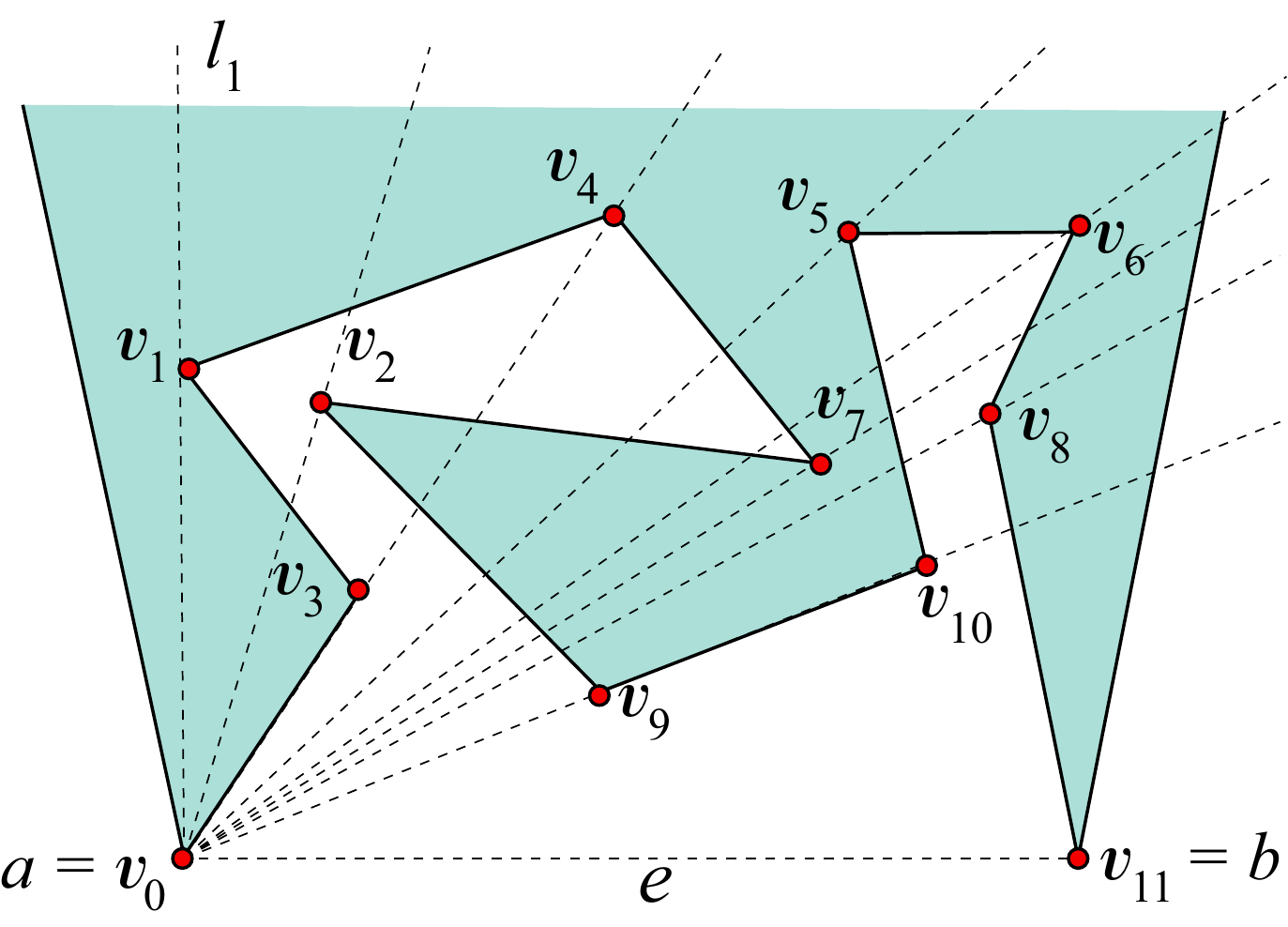} &
\includegraphics[width=0.32\linewidth]{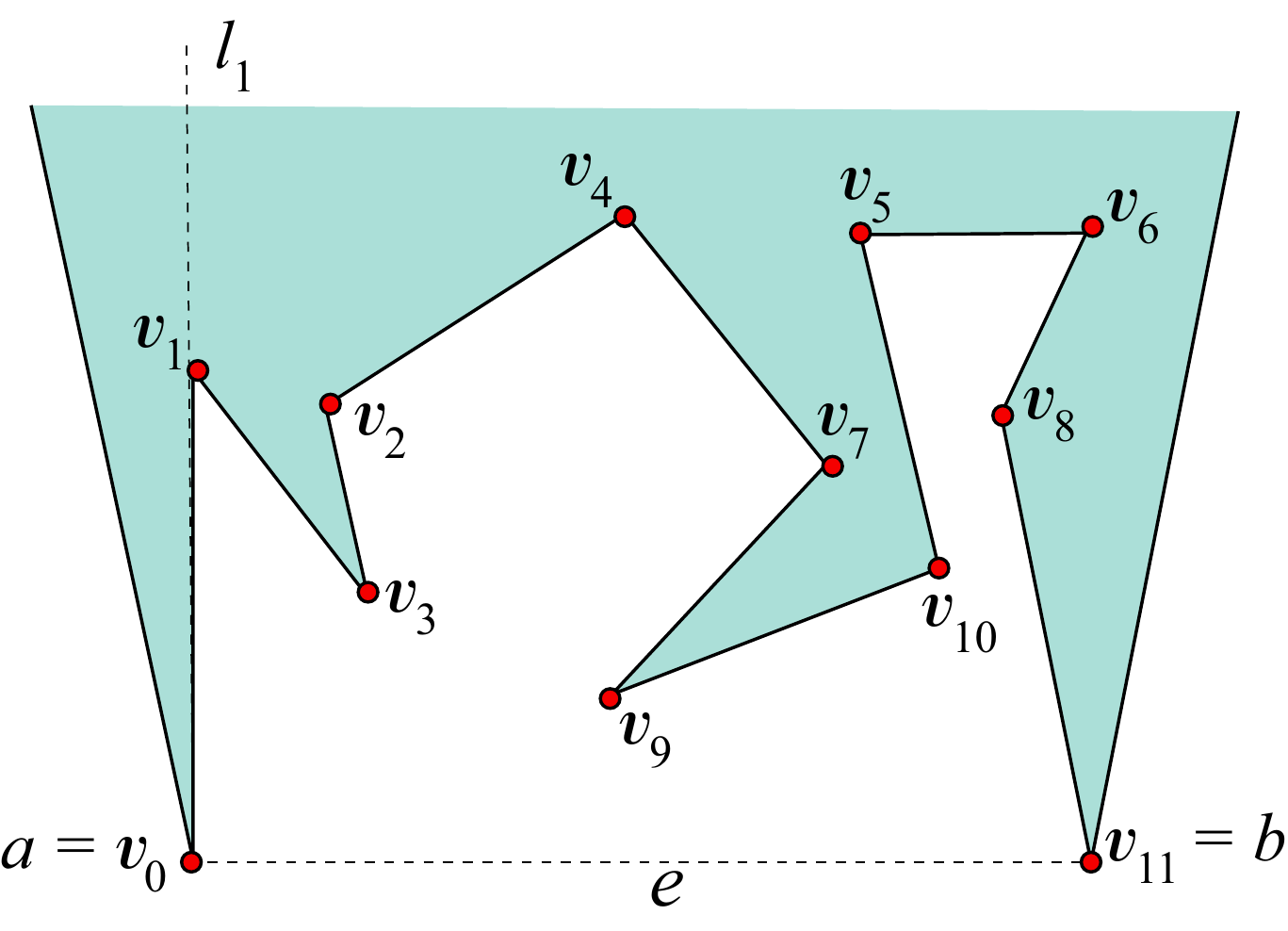} &
\includegraphics[width=0.32\linewidth]{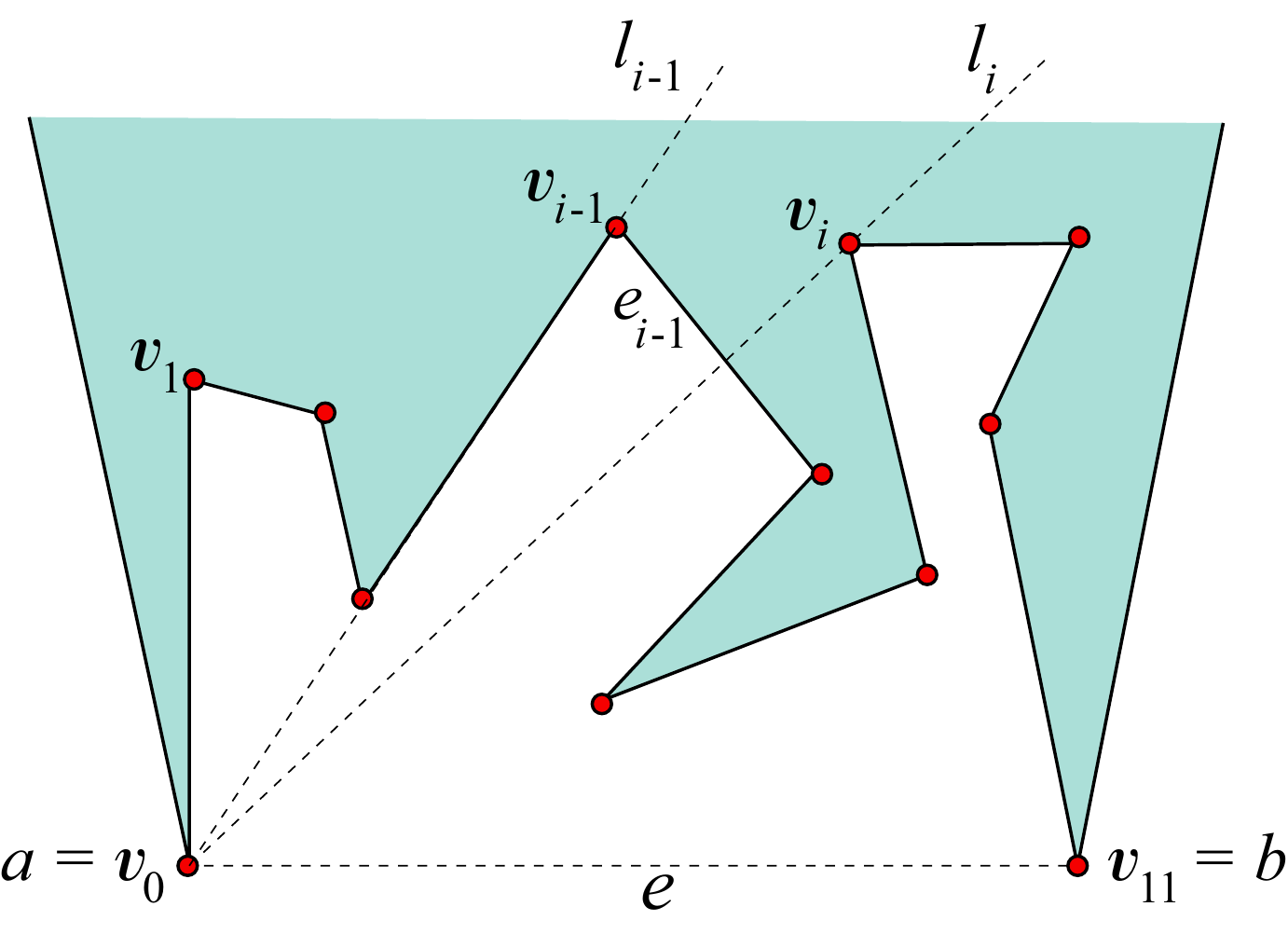}
\\
(a) & (b) & (c)
\end{tabular}
\caption{{\sc Canonical Polygonization}: \st$(v_i, x)$ is always
possible. (a) Base case ($i = 1$): $v_1$ sees $v_0$ (b) After
iteration 1, $v_0$ and $v_1$ are consecutive along the pocket
boundary (c) $v_i$ sees $x$.} \figlab{canonical.proof.firstfig}
\end{figure}

To prove the inductive step, suppose that the lemma holds for
iterations $1,\ldots,i-1$.  Note that the existence of the edge
$e_{i-1}$ selected in Step 2 of the algorithm follows immediately
from the fact that $v_0, v_1, \ldots v_{i-1}$ are consecutive along
the pocket boundary (cf. inductive hypothesis).
If $e_{i-1}$ is identical to $v_{i-1}v_i$, there is nothing to
prove. So assume that $e_{i-1}$ and $v_{i-1}v_i$ are distinct. We
now show that $v_i$ sees $e_{i-1}$, so that \st$(e_{i-1}, v_i)$ is
possible.

First observe that the wedge bounded by $\ell_{i-1}$ and $\ell_i$ is
either degenerate (if $v_0, v_{i-1}, v_i$ are collinear), or is
empty of any pocket points (since $v_i$ follows $v_{i-1}$ in the cw
sorted order). In the former case, $v_i$ sees $v_{i-1}$ and
$e_{i-1}$. In the latter case, $e_{i-1}$ must intersect $\ell_i$
(cf. Fig.~\figref{canonical.proof.firstfig}c). In either case, $v_i$
sees $e_{i-1}$ and hence \st$(e_{i-1},v_i)$ is 
possible. This
along with the induction hypothesis implies that at the end of
stretch operation, vertices $v_0, v_1, \ldots v_i$ are consecutive
along the pocket boundary.

Next we show by contradiction that the twang cascade of the forward
move involves only vertices $v_{i+1},\ldots, v_k$, so that
$v_0, v_1, \ldots v_i$ remain consecutive along the pocket boundary.
Suppose the claim is false. For ease of presentation, define
rank($v_i$) = $i$. Let $y$ be the first vertex with rank($y$) $\le
i$ to get into double contact. Clearly $y$ cannot coincide with $a$,
since $a$ is a hull vertex and cannot get into double contact. Let
\tw$(qrs)$ be the twang that created the double contact at $y$. Note
that at the time of \tw$(qrs)$, vertices $v_0, v_1, \ldots v_i$ are
consecutive along the pocket boundary, since none of these vertices
was in double contact prior to $y$ (by choice of $y$) and therefore
could not have twanged. Since \tw$(qrs)$ creates the double contact
at $y$, $y\in \triangle qrs$ and lies on $\sp(qrs)$. We also have
that rank($r$) $> i$, by our choice of $y$.


Two cases are possible: (i) $y$ lies strictly to the left of
$\ell_i$, 
and (ii) $y$ lies on $\ell_i$ 
In either case, since
$y\in \triangle qrs$ and $r$ lies on or to the right of $\ell_i$, it
must be that $\min\{$rank($q$), rank($s$)$\} <$ rank($y$). Suppose
w.l.o.g that rank($q$) $<$ rank($y$). Thus we have that rank($q$)
$<$ rank($y$) $\le$ rank($v_i$) $<$ rank($r$). In other words
rank($r$) $-$ rank($q$) $\ge 2$, but since $q \in \{v_0, v_1, \ldots
v_{i-1}\}$ and $qr$ is an edge of the pocket, it must be that
rank($r$) $-$ rank($q$) $=1$ (since $v_0, v_1, \ldots v_i$ are
consecutive along the pocket boundary). Thus we have reached a
contradiction. This completes the induction step.
\end{pf}

\section*{Appendix~2: Twang Cascade Lower Bound}
Fig.~\figref{double.twang} shows an example in which one point $v$
twangs twice in a cascade,
which gives a hint at the complex changes that can occur during
a cascade.
We will return to this example in Appendix~3 below.

\begin{figure}[htbp]
\centering
\includegraphics[width=\linewidth]{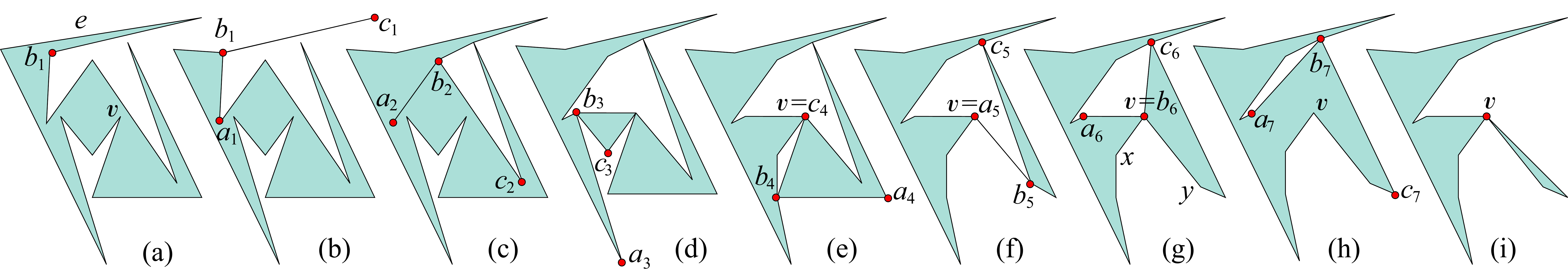}
\vspace{-1.7em}\caption{Point $v$ twangs twice: (a) Initial $P$ (b)
After  \st$(e, b_1)$ (c--i) After \tw$(a_ib_ic_i)$, $i = 1 \ldots 7$
(i) $v$ in double contact a second time, and can
twang a second time. } \figlab{double.twang}
\end{figure}

Figure~\figref{quadratic} displays an example in which 
$\Omega(n)$ points each
twang $\Omega(n)$ times in one cascade, providing an
$\Omega(n^2)$ lower bound on the length of a cascade.
The cascade is initiated by \st$(e,v)$ followed by \tw$(v)$.
From then on \tw$(a_i,b_i,c_i)$ twangs in a cycle.
Each such twang alters the path to $\sp(a_i,c_i)$,
which wraps around $b_{i+1}$.  In the next pass through the cycle,
\tw$(a_i,b_{i+1},c_i)$ occurs.
This continues just twice in this figure,
but in general the number of cycles is the number of
$b_j$ vertices inside each $\triangle a_i b_i c_i$.
\begin{figure}[htbp]
\begin{minipage}[b]{0.47\linewidth}
\centering
\includegraphics[width=0.95\linewidth]{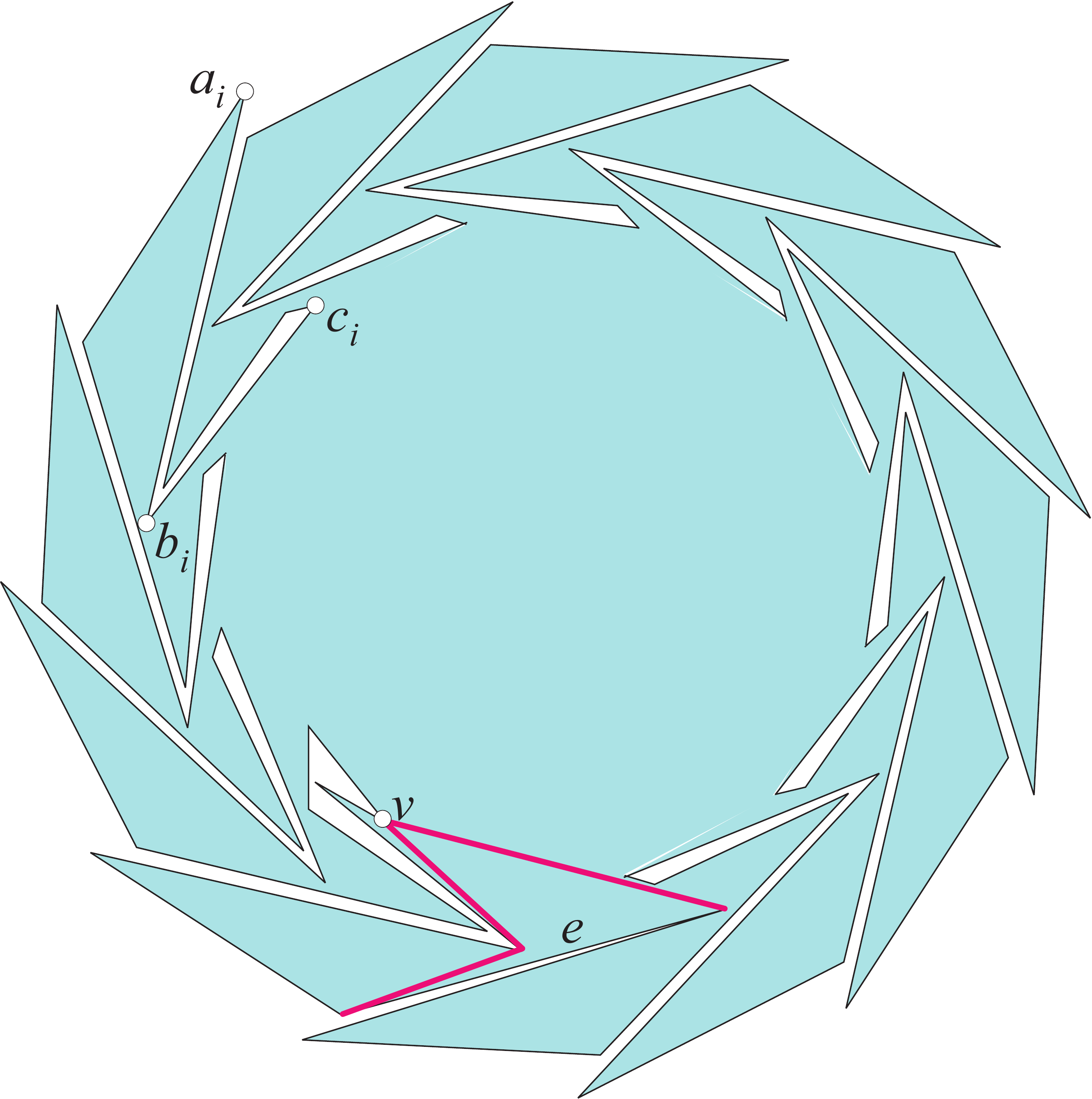}
\caption{\st$(e,v)$ followed by \tw$(v)$ initiates a 
quadratic-length twang cascade.}
\figlab{quadratic}
\end{minipage}%
\hspace{0.06\linewidth}%
\begin{minipage}[b]{0.47\linewidth}
\centering
\includegraphics[width=0.95\linewidth]{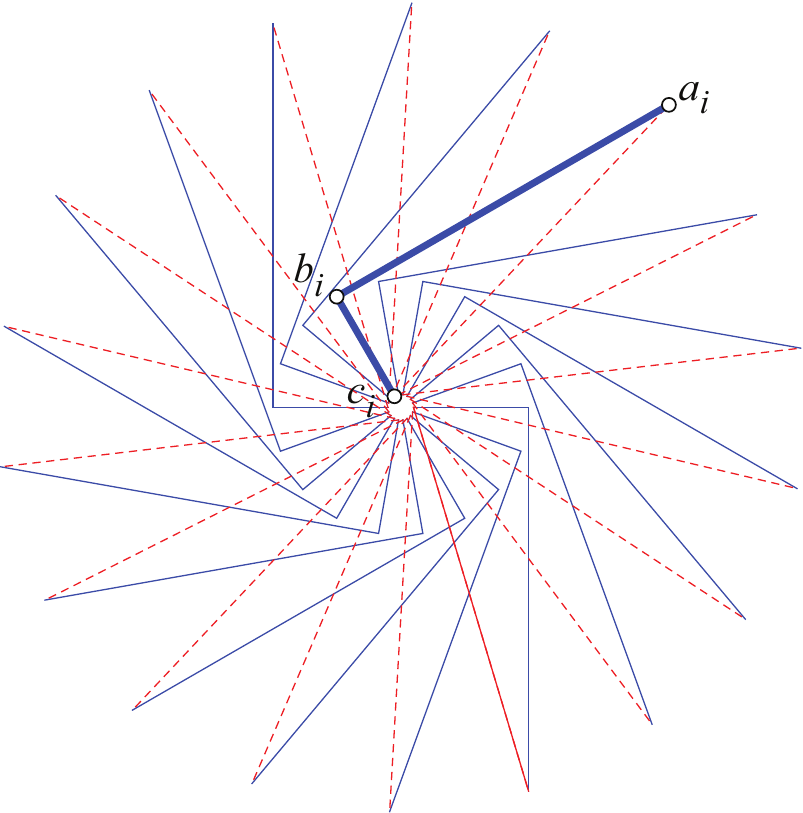}
\caption{It is possible for each triangle $\triangle a_i b_i c_i$
to enclose $\Omega(n)$ vertices $b_j$.
}
\figlab{pinwheel_quadratic}
\end{minipage}
\end{figure}
Figure~\figref{pinwheel_quadratic} shows that this number can be $\Omega(n)$.
Here the $b_j$ vertices are evenly spaced on a circle,
and $\angle (a_i,b_i,c_i) = 90^\circ$.
As the length $|a_i b_i|$ grows longer, the fraction of points
inside $\triangle a_i b_i c_i$ approaches $n/4$.

\section*{Appendix~3: Twang Cascade Upper Bound}
Figs.~\figref{twang.cascade},~\figref{double.twang}, and~\figref{quadratic}
support an
intuition that a twang cascade simplifies pocket entanglements in
some sense.
We capture this notion combinatorially in the \emph{pocket hierarchy tree} 
$T_P$ for a polygonization $P$,
a unique representation of the nesting of
pockets and subpockets of $P$ in a tree.
Each pocket has a \emph{level} $k$, with $P$ itself level 0,
the pockets
described in Section~\secref{Pockets} as level-$1$ pockets,
and its subpockets at level~$2$, and so on.
We will call all of these \emph{pockets}, and use
a superscript to distinguish the level $k$ and subscripts
to distinguish among the pockets at the same level:
$P^k_i$.

Each node of $T_P$
is the list of vertices in $\hull(P^k_i)$.
If $P^k_i$ is convex, it has no children.
Otherwise, each edge $ab$ of $\hull(P^k_i)$
which is not an edge of the polygonization $P$ is
a pocket lid for the vertices of $P$ from $a$ to $b$.
Fig.~\figref{subpockets} illustrates the hierarchy.
Note that every vertex $v \in P$ is on the hull of one
or more pockets of $T_P$.
\begin{figure}[htbp]
\centering
\includegraphics[height=0.96\textheight]{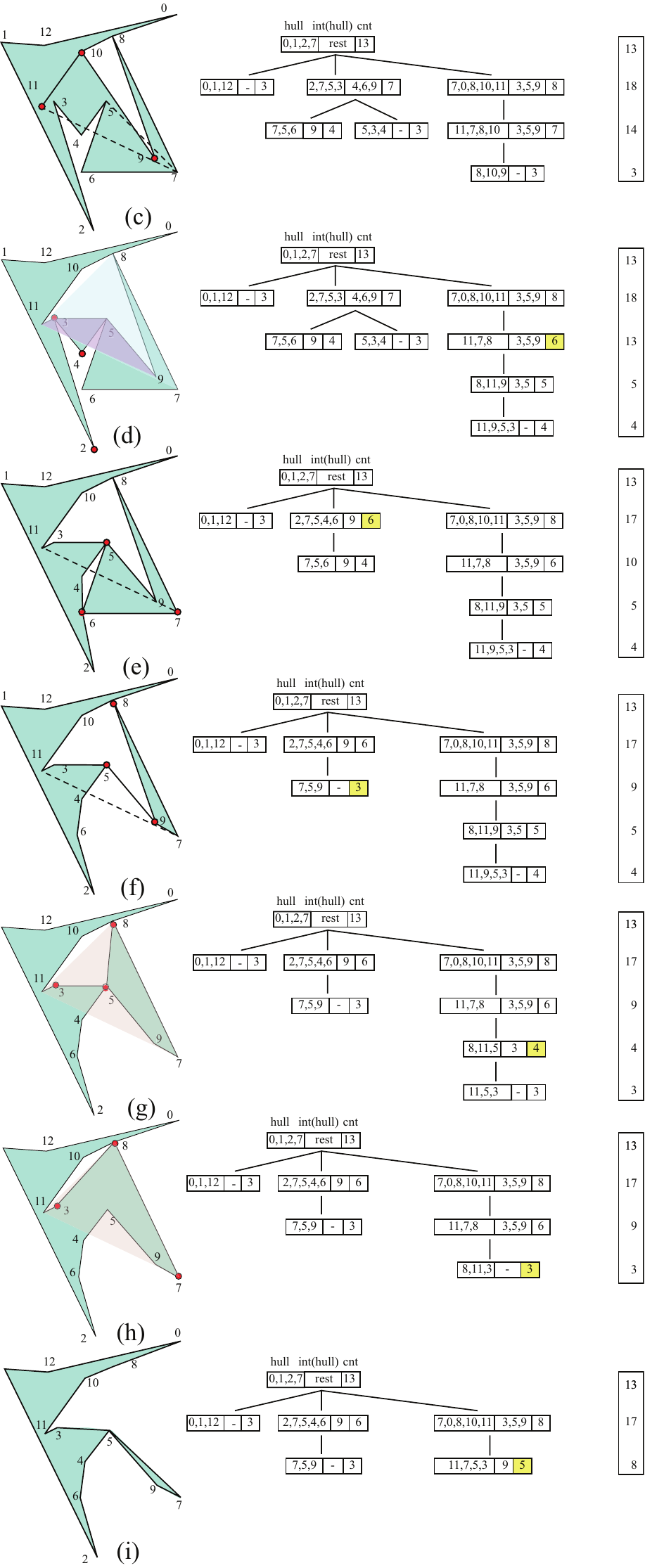}
\caption{Pocket hierarchy for cascade illustrated
in Fig.~\protect\figref{double.twang}.
Middle column displays the pocket tree
(each node showing:
points on hull, points interior, $S()$ pocket count), 
right column the pocket vector transposed.}
\figlab{subpockets}
\end{figure}

For each pocket 
$P^k_i$,
we define a \emph{pocket count} $S(P^k_i)$
to be the number of points on or in
$\hull(P^k_i)$.
For example, in 
Fig.~\figref{subpockets}c,
the pocket with lid $(7,0)$ contains points
$\{7,0,8,10,11,3,5,9\}$ and so has pocket count $8$.
Note that we count points $8$ and $10$ only once even though
they are in double contact.
Finally, we define the \emph{pocket vector} $V$ for
$P$ to list the sum of all pocket counts for
all pockets at each level:
$V= {<} V_1,V_2,V_3, \ldots {>}$,
$V_{k} = \sum_i  S(P^k_i)$.
The pocket vectors in Fig.~\figref{subpockets} are:

\begin{center}
\begin{tabular}{|c|l|}
\hline
(c)  &   ${<} 13,18,14,3 {>}$ \\
(d)  &   ${<} 13,18,13,5,4 {>}$ \\
(e)  &   ${<} 13,17,10,5,4 {>}$ \\
(f)  &   ${<} 13,17, 9,5,4 {>}$ \\
(g)  &   ${<} 13,17, 9,4,3 {>}$ \\
(h)  &   ${<} 13,17, 9,3 {>}$ \\
(i)  &   ${<} 13,17, 8 {>}$ \\
\hline
\end{tabular}
\end{center}

Let $V= {<} V_1,V_2,V_3, \ldots {>}$ and $W={<}W_1,W_2,W_3, \ldots {>}$ be two pocket vectors.
We define a lexicographic ordering relation on them:
$V < W$
iff
$V_1=W_1, V_2=W_2, \ldots, V_k=W_k, V_{k+1} < W_{k+1}$.
The steady odometer-like decrementing of the pocket vector
evident in the above example
is the sense in which a twang cascade simplifies pocket structure:

\begin{theorem}
A twang \tw$(abc)$ always strictly reduces the pocket vector.
\end{theorem}
\begin{pf}
Assume $P^k_i$ is the highest pocket in the hierarchy for which 
vertex $b$ is on $\hull(P^k_i)$. 
We know at least one such pocket exists.   

\begin{description}
\item[Case 1:]
$b$ is a lid vertex of pocket $P^k_i$.

We show by contradiction that this case is impossible: Since $b$ is a lid 
vertex of pocket $P^k_i$, it must be a vertex on the hull of the parent of 
$P^k_i$. But this contradicts our assumption that $P^k_i$ is the highest 
pocket having $b$ on its hull. (If $b$ has no parent because it is a level-1 
pocket, then it must be on the hull of the entire polygon, which is a 
contradiction since such vertices are never twanged.)

For example, in Fig.~\figref{subpockets}c,
let $b=10$. Then $b$
is a lid vertex of pocket $\{10,8,9\}$ at level 3, 
but it is also a non-lid hull vertex of pocket $\{11,7,8,10\}$ at level 2.

\item[Case 2:]
$b$ is not a lid vertex of pocket $P^k_i$.

This case implies that $a$, $b$, and $c$ are all part of pocket $P^k_i$'s 
boundary, as are edges $ab$ and $bc$. Because a twang is performed at $b$, we 
know $b$ is a corner point of hull($P^k_i$). Thus after the twang, $\hull(P^k_i$) 
loses this corner point and $S(P^k_i)$ goes down by one.

Now we show that the pocket counts for pockets at levels $< k$ do not change 
as a result of \tw$(abc)$, and in addition no other pocket besides $P^k_i$ 
at level $k$ changes. First, observe that in addition to $P^k_i$, the only 
pockets that may include edges $ab$ and $bc$ are $P^k_i$'s ancestors and 
descendents in the hierarchy. Therefore, \tw$(abc)$ affects only these 
pockets and no others.  Of these pockets, only changes in $P^k_i$'s 
ancestors could increase the pocket vector. We show now that the counts for the 
ancestors do not changed. First observe that only the lid vertices of a 
pocket are hull vertices of its parent. Since $b$ is not a lid vertex, 
\tw$(abc)$ cannot change the hull of its parent, and similarly cannot 
change the hull of any of its ancestors. If the hulls do not change, then 
the point counts for the pockets do not change.
\end{description}
\end{pf}

\begin{corollary}
A twang cascade can have at most $O(n^n)$ steps.
\corlab{odometer}
\end{corollary}
\begin{pf}
Let $m$ be the maximum length of a pocket vector:
$m=n/3 = O(n)$.
In the worst case, a decrement of $V_k$ by 1 is followed
by $V_k,\ldots,V_m$ each being (somehow) reset to $n$ and counting down
through their full range.
If this happens for every decrement, then
the ``odometer'' counts through all $m^n$ distinct possible pocket vectors.
\end{pf}

Note particularly the change in the pocket vector from 
Fig.~\figref{subpockets}c to~d: although $V_2$ decrements
from $14$ to $13$, $V_3$ increases from $3$ to $5$
(and, in addition, $T_P$ grows in depth).
It is this phenomenon that makes establishing a better bound
via the pocket hierarchy problematical.

\end{document}